%% file: paper.tex
\documentstyle[12pt,epsfig,color]{article}
\def\baselinestretch{1.3}
\advance\textheight by \topskip
\newcommand{\comment}[1]{}

\def\beq{\begin{equation}}
\def\eeq{\end{equation}}
\def\beqn{\begin{eqnarray}}
\def\eeqn{\end{eqnarray}}

\begin{document}
 \tolerance=100000
 \input{definition.tex}
 %#############################
\vspace*{\fill}
\vspace{-1.5in}
\begin{flushright}
%{July,2005}\\
%\today \\
%{\tt hep-ph/0512011}\\
{\tt IISER/HEP/05/10}
\end{flushright}
\begin{center}
{\Large \bf
SUSY darkmatter at the LHC - 7 TeV
}
  \vglue 0.4cm
  Nabanita Bhattacharyya$^{(a)}$\footnote{nabanita@iiserkol.ac.in},
  Amitava Datta$^{(a)}$\footnote{adatta@iiserkol.ac.in} and
  Sujoy Poddar$^{(b)}$\footnote{sujoy$\_$phy@iiserkol.ac.in}
      \vglue 0.1cm
          {\it $^{(a)}$
	  Indian Institute of Science Education and Research, Kolkata, \\
%	  P.O. Mohanpur, PIN 741246, West Bengal,India.\\
          Mohanpur Campus, PO: BCKV Campus Main Office,\\
          Mohanpur - 741252, Nadia, West Bengal.\\
	  \it $^{(b)}$
	  Netaji Nagar Day College,\\
	  170/436, N.S.C. Bose Road, Kolkata - 700 092, India.
	  \\}
	  \end{center}
	  \vspace{.1cm}

%\begin{document}
%\thispagestyle{empty}
\begin{abstract}
We have analysed the early LHC  signatures  of the minimal
supergravity (mSUGRA) model. Our 
emphasis is on the 7 - $TeV$ run corresponding to an  integrated luminosity of 
$\sim 1.0 ~fb^{-1}$ although we have also  discussed briefly the prospects at 
LHC-10 $TeV$. 
We focus on the parameter space yielding relatively light squark and 
gluinos 
consistent with the darkmatter relic density data and the LEP bounds on the 
lightest
Higgs scalar mass. This parameter space is only allowed for non-vanishing 
trilinear soft breaking term 
$A_0$. A significant region of the parameter space with large to 
moderate negative values of $A_0$  consistent with the stability of the scalar 
potential and 
relic density production 
via neutralino  annihilation and/or neutralino - stau coannihilation yields 
observable
signal via the jets + missing transverse energy channel. The one lepton + jets +
missing energy signal is also viable over a smaller but non-trivial parameter 
space. The ratio of the size of the two signals - free from theoretical 
uncertainties -
may distinguish between different relic density generating mechanisms.      
With efficient $\tau$-tagging facilities at 7 $TeV$ the discriminating power may 
increase significantly.
We also comment on other dark matter relic density allowed
mSUGRA scenarios and variants there of
in the context of LHC-7 $TeV$.

\end{abstract}

%\maketitle

\section{Introduction}

%Low energy LHC runs at 7 Tev and 10 TeV.

The attention of the high energy physics community has been focussed on
the prospects of new physics search at the CERN Large Hadron Collider
(LHC) \cite{review}. 
It is gratifying to note that the proton -proton collisions with
stable beams is now operational at an energy  ($\sqrt{s}$ = 7 $TeV$)
never attained by any
accelerator before. It was of course expected that before operating
at the maximum attainable energy of $\sqrt{s}$ = 14 $TeV$, the warming 
up
exercises will begin with low energy and low luminosity runs. However,
due to the unfortunate accident which delayed the programme for about a
year,
the LHC operations has taken some unexpected twists and turns. For
example, until a couple of months ago it was planned that a substantial
amount of data will be delivered at $\sqrt{s}$ = 10 $TeV$ and several
analyses on new physics research have been published \cite{tev10}.
%({\bf integrated luminosity range (0.1 - 0.2) $\ifb$}).
However, a very recent revised decision has opted for data taking for
the next 18 - 24 months at $\sqrt{s}$ = 7 $TeV$ the anticipated integrated
luminosity being of the order of 1 $fb^{-1}$.  In some very recent 
analyses
\cite{7tev}  supersymmetry (SUSY) searches at $\sqrt s = 7~TeV$ 
%for different luminosities ranging from (0.1 -2) $\ifb$ 
have also been performed.  
Although this is a
temporary set back for the new physics search programme, it is
worthwhile to check what can be achieved at $\sqrt{s}$ = 7 $TeV$. In this
paper we concentrate on SUSY \cite{susy} 
search within the framework
of the simplest gravity mediated supersymmetry breaking model - the
minimal supergravity \cite{msugra}(mSUGRA) model - with the parameter
space constrained by the dark matter relic density data \cite{wmap}.

% Supersymmetry discovery potential of the LHC at s**(1/2) = 10 TeV
%and 14-TeV without and with
%missing E(T).
%Howard Baer, (Oklahoma U.) , Vernon Barger, (Wisconsin U., Madison) ,
%Andre Lessa, (Oklahoma U.) ,
%Xerxes Tata, (Hawaii U.) . UH-511-1142-09, Jul 2009. 34pp.
%Published in JHEP 0909:063,2009.
%e-Print: arXiv:0907.1922 [hep-ph]

In the mSUGRA model
there are only five free parameters \cite{msugra}
namely a common scalar mass ($m_0$),
a common gaugino mass ($m_{1/2}$), $tan \beta$, $A_0$ and sign of $\mu$.
Here $tan \beta$ is the ratio of the vacuum expectation values of the
two neutral Higgs bosons in the model, $A_0$ is the trilinear soft 
breaking term
and $\mu$ is the higgisino mass parameter; the magnitude of
$\mu$ is fixed by the radiative electroweak symmetry breaking (EWSB) 
condition \cite{ewsb}.

It is obvious that the strongly interacting squarks and gluinos 
are the species most
likely  to show up under the LHC microscope provided they
are relatively light with masses just beyond the reach of Tevatron Run 
II or  little more. 
%In mSUGRA scenario 
We
recall that the current lower limits \cite{datateva} 
obtained by the CDF collaboration  are
392 $GeV$ for squark masses assuming squark and gluino masses to be 
approximately equal 
and 280 $GeV$ for gluino masses with $m_{\wt q} \gsim 600 ~GeV$,  
in the mSUGRA scenario with $A_0 =0$,
$\mu > 0$ and $\tan \beta =5$.  
Similar limits from the  D$\dzero$ collaboration \cite{d02plb08} 
on the gluino mass are
379 $GeV$ and 307 $GeV$ respectively within the
framework of mSUGRA.

The low $m_0$ -
$\mhalf$ region of the parameter space yields light squarks and gluinos.
This parameter space also predicts light sleptons which facilitates
neutralino pair annihilation (or bulk annihilation) 
\cite{recentSUSYDMreview,bulk} 
via slepton exchange leading to
dark matter (DM) relic density consistent with data.

On the other hand a significant region of the above parameter space is
strongly disfavoured by the lower bound on the lightest Higgs boson mass
($m_h > 114.5 ~GeV$) from Large Electron Positron 
(LEP) \cite{higgsbound}. This leads to the
belief that bulk annihilation cannot be an important relic density
producing mechanism.

It has been  pointed out \cite{debottam} that the above conclusion is
an
artifact of the ad hoc assumption that the trilinear soft breaking
parameter $A_0 =$ 0, which has been invoked by most of the existing
analyses \cite{dmanalyses}. On the other had for a reasonably large 
negative value of
$A_0$, which does not violate the charge colour breaking condition
\cite{ccb}, the
LEP bound on $m_h$ can be satisfied with smaller values of $m_0$
and $m_{\half}$ which are excluded for $A_0 =$ 0. In the presence of
large negative $A_0$, two  interesting zones (referred to as I and II) 
of the parameter
space consistent with both Wilkinson Microwave Anisotropy Probe (WMAP) 
\cite{wmap} and LEP data \cite{higgsbound} open up:

I) In this zone with large negative $A_0$ bulk annihilation 
\cite{bulk} is the dominant DM producing
mechanism although LSP-$\stau$ coannihilation \cite{coann}
has a significant presence.

II) In this region in contrast to Region I, LSP-$\stau$ coannihilation
dominates while bulk
annihilation produces a moderate but non-negligible fraction of the
relic density.
Even if  LSP-$\stau$ coannihilation is the only significant
DM producing mechanism, it may occur for smaller values of
$m_{\half}$ compared to the  $|A_0|$ = 0 case. This allows lighter
squarks and gluinos.

Several examples of the above two zones are contained in Figs.
1 - 4 of \cite{debottam}.
In order to study the characteristic LHC signals corresponding
to different relic density producing mechanism 
the representative points A (zone I) and B (zone II; see Table 1 
reproduced from \cite{debottam}) are introduced. In order to compare 
with the 
signatures corresponding to the conventional parameter space with
$A_0$ = 0,  a third point C is also considered. This point 
corresponds to the minimal value of $\mhalf$ for $m_0 = 120~ GeV$
consistent with the WMAP data.
The expected event characteristics at the LHC 
for different representative points will be summarized in 
section 2.

It was further shown in \cite{debottam} that the lepton flavour
($e$, $\mu$, $\tau$)
content of the final states arising from squark-gluino events at the
LHC, will be
very different in the above cases. Tagging of $b$-jets may further help
to discriminate among scenarios. 
The observability of the these signals at the LHC ($\sqrt{s}$ = 14 $TeV$)
has been  illustrated \cite{debottam,nabanita} by 
Pythia \cite{pythia} based analyses. The  $\tau$-jet and $b$-jet tagging 
efficiencies
reported by the CMS collaboration \cite{cms} were used (see
Table IX of \cite{nabanita}). However, the
flavour tagging efficiencies  at $\sqrt{s}$ = 7 $TeV$ are not yet known. 
Hence
the flavour tagging  can not be readily employed for assessing the
physics potential of the low energy runs.

It is indeed gratifying to note that generic SUSY signals consisting of
$m$-leptons
+ $n$-jets + $\etslash$  without any flavour tagging also differ 
dramatically in the above
three scenarios  \cite{nabanita} (see Table IV). The main purpose of 
this paper is to scrutinize
different relic density producing  mechanisms and
check
the feasibility of distinguishing among them by using these generic
signals at LHC runs at energies lower than  the maximum 
attainable one. It should be stressed that the  parameter space
studied in this paper and its physical significance in the context of 
SUSY dark matter is totally different from the other studies 
\cite{7tev}.  

At the  focus of our attention is the physics at $\sqrt{s}$
= 7 $TeV$ (section 3). However, the strategy for future runs at the LHC 
will depend on the
performance of the experiments at 7 $TeV$. Another
round of  experiments at
an energy lower than the maximum attainable energy is cannot
not be ruled out  as yet.
We have, therefore, briefly discussed the signatures of the
above scenarios at $\sqrt{s}$ = 10 $TeV$, the  proposed energy for the
preliminary runs until very recently (section 4).

Although we have focussed on the benchmark points A, B
and C, we have also considered  several other points consistent with 
the relic 
density data. These points belong to  different regions of the parameter 
space yielding distinct sparticle spectra and collider signatures. The 
points A, B and C are chosen with $tan \beta$
= 10. In this analysis the  allowed points at higher 
as well as lower values of $tan \beta$ have been considered   
to make the study more comprehensive (section 3). Using reasonable
guesses about the efficiency, we have also considered the prospect of
observing $\tau$-jet tagged
signals at the low energy runs at  7 - $TeV$.  We have also taken this
opportunity to improve  some of our earlier background estimates in
\cite{debottam,nabanita}.

Other authors have also proposed schemes for testing the origin of dark 
matter production at the LHC \cite{liu}.
%----------------------------------------------------------

\section{The allowed mSUGRA parameter space for non zero trilinear
soft breaking terms}
%----------------------------------------------------------

%In the simplest gravity mediated SUSY breaking model -
%the minimum super gravity (mSUGRA) model
%there are only five free parameters \cite{msugra}
%namely a common scalar mass ($m_0$),
%a common gaugino mass ($m_{1/2}$), $tan \beta$, $A_0$ and sign of $\mu$.
%Here $tan \beta$ is the ratio of the vacuum expectation values of the
%two Higgs bosons in the model, $A_0$ is the trilinear soft breaking term
%and $\mu$ is the higgisino mass parameter; the magnitude of
%$\mu$ is fixed by the radiative electroweak symmetry breaking condition
%\cite{ewsb}.

In \cite{debottam} signals at the  LHC corresponding to the WMAP 
allowed regions of the
parameter space with non-zero $A_0$ were studied at 14 $TeV$.
The results were
compared and contrasted with the expectations from the well publicized
conventional $\stauone$-coannihilation 
\cite{recentSUSYDMreview,coann}
scenario with $A_0 = 0$
by introducing three benchmark  points A, B and C.
%(Table I of \cite{debottam}, which has  been 
%for a ready reference). 
The corresponding
mSUGRA parameters are reproduced in 
Table 1 for ready reference.

%The scenario C corresponds to the
%smallest $\mhalf$
%which is consistent with the Higgs mass bound and the WMAP data.

\begin{table}[!htb]
\begin{center}\

\begin{tabular}{|c|c|c|c|}
       \hline
       mSUGRA &A&B&C\\
       parameters & &&\\
       \hline
       $m_0$ &120.0  &120.0 &120.0\\
       \hline
       $m_{1/2}$ &300.0  &350.0 &500.0\\
       \hline
       $A_0$ &-930.0  &-930.0 &0.0\\
       \hline
       $\tan\beta$ &10.0  &10.0 &10.0\\
       \hline
       $sign(\mu)$ &1.0  &1.0 &1.0\\
       \hline

       \end{tabular}
       \end{center}
          \caption{Three bench mark scenarios introduced in \cite{debottam}.
	  All parameters with dimension of mass are in $GeV$.}
	  \end{table}

 The sparticle spectra in the three scenarios can be found in Table II of
 \cite{debottam}. The total cross section of squark gluino events 
 decreases significantly as we go from A to C, which is easily 
understood from the respective sparticle spectra. In scenario A the 
lighter top squark-antisquark ($\lstop \lstop^*$) pair production enhances the 
total cross section significantly. This is a direct consequence of large 
negative $A_0$. This trend is also seen at 7 - $TeV$ (see Table 2).

  The signals at the LHC are governed by the cascade decays of the
  sparticles. In all three cases the gluinos being heavier than all squarks, decay
  into quark-squark pairs (Table III of \cite{debottam}). Decays into $t-\lstop$ pairs 
  dominate in scenarios A and B (Table IV of \cite{debottam}), as 
  the third generation squarks are relatively light 
  due to the renormalization group evolution and large $|A_0|$.
  
  The squarks in general decay into the corresponding lighter quarks and
  an appropriate electroweak gaugino. 
  The decay of each third
  generation squark inevitably contains a bottom ($b$) quark. 
  This is the origin of the
  large fraction of final states with $b$-jets as noted in \cite{debottam}.
  In scenario C the fraction of third generation squarks in gluino decay
  is relatively small and the above effect is suppressed.
  
The decay properties of the lighter chargino ($ \chonepm$)and the second lightest
  neutralino ($\lsptwo $) which, in addition to the LSP, are often present in squark-gluino
  decay chains control the lepton content of the final sates to
  a large extent. As a  direct consequence of the presence of the light
  sleptons,  these two
  unstable gauginos decay almost exclusively into leptonic channels via two body
  modes in all three scenarios.
  In scenario A the lighter chargino decays into R-type sleptons
  with a large BR This results in a very large fraction of final states
  containing the $\stauone$ which is always lighter than the other sleptons and
  eventually decays
  into a $\tau$-LSP  pair (see Table V of \cite{debottam} and Table II 
  of \cite{nabanita}).
  The $\lsptwo$ decays primarily into $\tau$-$\stauone$ pair contributing
  further to the $\tau$ dominance in the final states.
  The scenario B has all the above
  features albeit to a lesser extent.
  In scenario C the $\chonepm$ decays into left slepton- neutrino pairs
or sneutrino -lepton pairs of all  generations with almost 
equal BR of sizable magnitudes
and lepton universality holds
to a very good approximation.

%----------------------------------------------------------

\section{The Signals at the LHC at 7 $TeV$}

In this analysis we have generated all squark-gluino events at $E_{CM} = 
7~TeV$ using Pythia \cite{pythia}. Initial and final state radiation, 
decay, hadronization, fragmentation and jet formation are implemented 
following the standard procedures in Pythia. The lowest order 
squark-gluino production cross-sections have been computed by CalcHEP 
\cite{calchep}. The corresponding cross-sections for the scenarios A,  
B and C are presented in Table \ref{xsec}. 
%---------------------------------------------------------------------- 

\begin{table}[!ht]
\begin{center}\

\begin{tabular}{|c|c|c|c|}
\hline
%     & \multicolumn{3}{c|}{}
& \multicolumn{3}{c|}{$\sigma(\pb)$}\\
\cline{2-4}
Process&A&B&C\\
\hline
$\wt g \wt g$            &   0.040  & 0.010 & 0.26 $\times 10^{-3}$\\
$\qL \wt g$              &   0.140  & 0.057 & 2.44 $\times 10^{-3}$\\
$\qR \wt g$              &   0.155  & 0.063 & 2.78 $\times 10^{-3}$\\
$\qL \qL$                &   0.113  & 0.057 & 5.62 $\times 10^{-3}$\\
%$\qL \qL^*$              &   0.108  & 0.029 & 0.0023\\
$\qR \qR$                &   0.100  & 0.051 & 5.49 $\times 10^{-3}$\\
%$\qR \qR^*$              &   0.095  & 0.047 & 0.0032\\
$\qL \qR$                &   0.059  & 0.055 & 2.08 $\times 10^{-3}$\\
%$\qL \qR^* + c.c$        &   0.149  & 0.067 & 0.0077\\
$\lstoppair$             &   0.928  & 0.162 & 0.047 $\times 10^{-3}$\\
%$\hstoppair$             &   0.02  & 0.008 & 0.0006\\
%$\lstop \hstopbar + c.c$ &   0.0002 & 0.0002 &6.8$\times 10^{-6}$\\
%$\lsbotpair$             &   0.047  & 0.017 &0.0008\\
%$\hsbotpair$             &   0.019  & 0.008 &0.0006\\
\hline
\hline
Total & 1.535&0.455 &0.018 \\
\hline
\end{tabular}
\end{center}
\caption{The production cross sections of all squark-gluino events
studied in this paper.}
\label{xsec}
\end{table}

%%%%%%%%%%%%%%%%%%%%%%%%%%%%%%%%%%%%%%%%%%%%%%%%%%%%%%%%%%%%%%%%%

We have used the toy calorimeter simulation (PYCELL) provided in Pythia 
with the following criteria:

\begin{itemize}
\item The calorimeter coverage is $\vert \eta \vert < 4.5$. The segmentation is given by $\Delta \eta \times \Delta \phi = 0.09 \times 0.09$ which
resembles a generic LHC detector.

\item A cone algorithm with $\Delta$ R$ = \sqrt {\Delta\eta^2 + \Delta\phi^2}= 0.5 $ has been used for jet finding.

\item E$^{\mathrm{jet}}_{\mathrm{T,min}} = 30 ~GeV$ and jets are ordered 
in E$\mathrm{_T}$.
\end{itemize}

%********
The stable leptons are selected according to the criterion :
\begin{itemize}

\item Leptons $(l=e,\mu)$ are selected with $P \mathrm{_T \ge 20}~ GeV$
 and $\vert\eta \vert < 2.5$. For lepton-jet isolation
 we require $\Delta R(l,j) > 0.5$. For the sake of simplicity
the detection efficiency of $e$ and $\mu$ are assumed to be $ 100 \%$.

 \end{itemize}
 The following cuts are implemented for background rejection :
 \begin{itemize}

 %\item Leptons $(l=e,\mu)$ with P$\mathrm{_T \le 60}~ GeV$ are rejected to
 %ensure the rejection of leptons coming from $\tau$ decay. (CUT 1)

 \item We have required two leading jets having $P_T^{j_1}> 100 ~GeV $
 and $P_T^{j_2}> 75 ~GeV$.(CUT 1)

 \item Events with missing transverse energy ($\etslash) < 350~ GeV$ are rejected. 
(CUT 2)

 \item Events with $M_{eff} < 750~GeV$ are rejected,
 where $M_{eff}= |\met| + \Sigma_{i}|P_T^{l_i}| + \Sigma_{i}|P_T^{j_i}|$
 ($l = e,\mu$ ). (CUT 3)

 \item Only events with jets having S$\mathrm{_T} > 0.2$, where
 S$\mathrm{_T}$ is a standard function of the eigenvalues of the
 transverse sphericity tensor, are accepted. (CUT 4)
 \end{itemize}

 These cuts are motivated by the analysis of generated squark-gluino events
 by CMS collaboration \cite{cms2} although we have relaxed some of them in 
 view of the reduced $\sqrt{s}$.
 We begin with the generic SUSY signals of the type
 $m$-$l$ + $n$-$j$ + $\etslash$ , where $l = e$ or $\mu $ and $j$ is any jet.
 For establishing these  generic signals Cut 1 - Cut 4  are adequate.

%%%%%%%%%%%%%%%%%%%%%%%%%%%%%%%%%%%%%%%%%%%%%%%%%%%55

\begin{figure}[tb]
\begin{center}
\includegraphics[width=\textwidth]{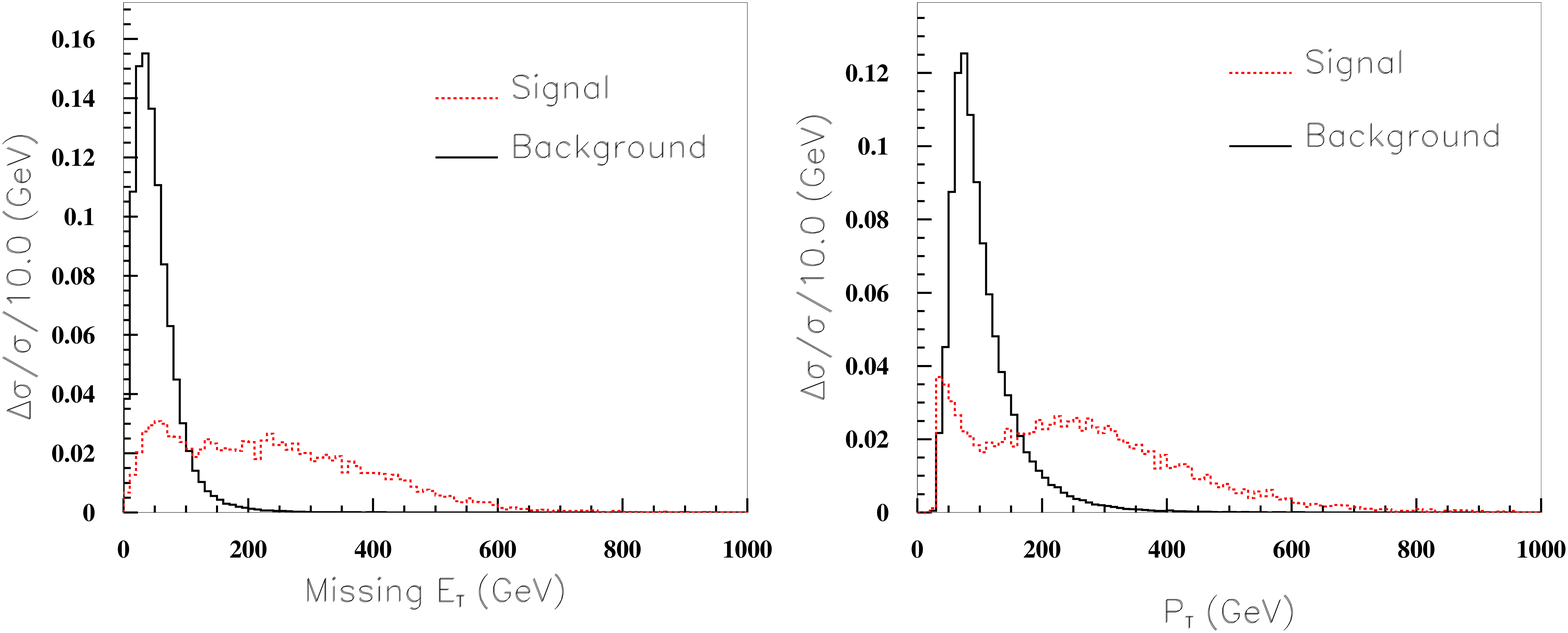}
\end{center}
\caption{The distributions (normalised to unity) 
of $\met$ (left) and $P_T^{j_1}$ (right)
for $0l$ events (before the selection cuts) for the signal (scenario A)
and the dominant backgrounds.}
\label{fig1}
\end{figure}

%%%%%%%%%%%%%%%%%%%%%%%%%%%%%%%%%%%%%%%%%%%%%%%%%%%55

 We have considered the backgrounds from  $t \bar t$, QCD events,
 $W$ + $n$-$jets$ and $Z$ + $n$-$jets$ events, where $W$ and $Z$ decays into all channels.

We have generated $t \bar t$ events using Pythia and the LO cross-section
has been taken from CalcHEP which is 68 $pb$.
For QCD processes generated by Pythia the contribution from the
$\hat p_T$ bin $ 400 ~GeV < \hat p_T < 1000 ~GeV$ has been considered
where $\hat p_T$ is defined in
the rest frame of the parton parton collision.
The cross-section for this bin is 227 $pb$. 
However, for other bins ($ 25 ~GeV < \hat p_T < 400 ~GeV$ and 
$1000 ~GeV < \hat p_T < 2000 ~GeV$), the background events are
negligible.

 For $W$ + $n$-$jets$ events we have generated events with $n=0,1$ and $2$ 
 at the parton level using ALPGEN (v 2.13) \cite{alpgen}.
 We have generated these events subjected to the condition that $P_T^j > 60~
 GeV$. These partonic events have been fed to Pythia for parton showering,
 hadronozation, fragmentation and decays etc.

 Similarly we have also generated  $Z$ + $n$-$jets$, events 
 with $n=0,1$ and $2$ using ALPGEN (v 2.13). 
 The partonic  jets have $P_T > 60 ~GeV$ and fed then
 to Pythia for further analysis.

In Fig. \ref{fig1} we have presented the normalised distributions of 
$\met$ (left panel) and $P_T^{j_1}$ (right panel) for $0l$ events. 
The  dominant backgrounds are from
$W$ + $n$-$jets$ and  $Z$ + $n$-$jets$ with $n=1$ and $2$. These 
distributions
motivate CUT 1 and CUT 2.

 For both the 1 gauge boson + $n$-jets
backgrounds the contribution from $n=0$ is negligible. The total background and
the individual contribution of each channel to it are in Table 4.
% In this analysis we have concentrated on generic signal type $0l$ and $1l$
% only where $l$ stands for $e$ or $\mu$. 
% For scenario A, number of $0 l$ and $1 l$ events for an integrated
% luminosity of $1 \ifb$ are 139 and 16.
% For scenario B, number of $0 l$ and $1 l$ events for an integrated
% luminosity of $1 \ifb$ are 80 and 12.
 We have computed the significance ($S \over \sqrt{B}$) for an integrated
 luminosity of $1 ~\ifb$ where $S$ is the number of signal events and $B$ 
 denotes the background events. For scenario A the significance for $0l$ and 
 $1l$ are 13.8 and 2.8 respectively (from Tables 3 and 4). 
 For scenario B (C) the corresponding numbers 
 are  7.9 (0.51) and 2.19 (3.5 $\times 10^{-4}$). 

 The sizable uncertainties in the counting rates at $\sqrt {s} =14 ~TeV$
 due to the choice of the parton density functions (PDF) and the QCD scale
 has already been  discussed  (see Table III of \cite{nabanita}). 
 It also follows 
from the same table that the ratio of two
 cross-sections with different $m$ and $n$ is remarkably stable with respect to 
 the above uncertainties. Thus to identify a particularly relic density 
 producing scenario unambiguously at least two measurements, for example the 
 number of $0l$ and $1l$ events, are essential. However, the numbers in the 
last paragraph 
\ref{signal} indicate that the statistics for $1l$ events may not be 
adequate  for $\lum = 1 ~\ifb$. With 
 a luminosity of 2 $\ifb$ the ratio $R = {\sigma_{0l} \over \sigma_{1l}}$
 can be measured with some confidence in scenario A. On the other hand if 
 no signal is seen even in the $0l$ channel, point A will be strongly 
 disfavoured in spite of the uncertainties.
 
 In order to make the scanning more comprehensive we have scanned 
 several points around A chosen from Fig. 1a) of \cite{debottam} 
 belonging to zone I defined in the introduction. For $\mhalf$ lower 
 than that of A, the cross section of $0l$ events is huge.
  For example, with
 $m_0=120, m_{1/2} =250, A_0=-930$ we find ${({S \over \sqrt{B}})}_{0l}$ 
 $\approx 38$ at $\lum = 1 ~\ifb$.  All mSUGRA mass parameters are 
 in $GeV$ unless  mentioned otherwise.
 Thus even the 7 - $TeV$ run can 
 probe bulk annihilation as a viable relic density producing mechanism
 for non-zero trilinear couplings or disfavour it. 
Similarly a large number of points in 
 zone II around point B were scanned and ${({S \over \sqrt{B}})}_{0l} \gsim 5$
 were obtained in most of the cases. However, a large part of zone II 
 will remain unexplored at $\sqrt{s} = 7 ~TeV$.

%%%%%%%%%%%%%%%%%%%%%%%%%%%%%%%%%%%%%%%%%%%%%%%%%%%%%%%%%%%%%%%%%%%%

\begin{table}[!ht]
\begin{center}\
\begin{tabular}{|c|c|c|c|c|c|c|}
%\hline
%&\multicolumn{3}{c|}{SIGNAL} & \multicolumn{3}{c|}{background}\\
%\cline{2-5}
\hline
& A & B & C& D &E & F  \\
\hline
$\sigma (\pb)$    & 1.535       & 0.465   & 0.018   & 3.2  & 11.3   & 10.3 \\
\hline 
%\hline
$ 0 l $       &0.1386        &0.0798    &0.0051   & 0.125  & 0.316  & 0.287\\
\hline
$ 1 l $       & 0.0161       & 0.0122  & 0.0020  & 0.0384  &0.026   & 0.033\\
\hline
$ 1\tau + X $ & 0.010      & 0.0051    & 0.0003 & 0.005  & 0.024     & 0.015\\
\hline
\end{tabular}
\end{center}
 \caption{The cross-sections (including efficiency) at $Q = \sqrt{\hat s}  $ 
 for signal events
with different $m$. For details of D, E and F see text.
}
\label{signal}
\end{table}

%%%%%%%%%%%%%%%%%%%%%%%%%%%%%%%%%%%%%%%%%%%%%%%%%%%%%%%%%%%%%%%%%%%%
\begin{table}[!ht]
\begin{center}\
\begin{tabular}{|c|c|c|c|c|c|c|c|}
%\hline
%&\multicolumn{3}{c|}{SIGNAL} & \multicolumn{3}{c|}{background}\\
%\cline{2-6}
\hline
& $t\bar t$ & $W + 1j$ & $W + 2j$& $Z+ 1j$&$Z+2j$ & QCD&Total \\
\hline
$\sigma (\pb)$    & 68       & 1836   &  446  & 652  & 122   & 227& Background\\
\hline 
\hline
$ 0 l $       & 0.00374    & 0.0129   & 0.0241  & 0.0196 &0.03686  &0.0043 &0.1015 \\
\hline
$ 1 l $       & 0.00129    & 0.0165   & 0.0147  & -- & -- &-- &0.0328 \\
\hline
$1 \tau + X$  & 0.00034    & 0.0013   & 0.0024  & --   & 0.00046   & 0.0015  & 0.006       \\
\hline
\end{tabular}
\end{center}
 \caption{The cross-sections (including efficiency) at $Q = \sqrt{\hat s}  $ for background process
with different $m$.
No entry in a particular column (-) means negligible background.
}
\label{bgnd}
\end{table}

Although ${({S \over \sqrt{B}})}_{1l}$ is less than 5 in scenarios A and 
B, larger values show up 
in several regions allowed by relic density data and LEP bound on $m_h$. 
At these points with different values of $tan \beta = 5,10$ and $20$ the 
cross sections (see Table
3 for illustrations) are significantly larger compared to A or B. 
As a result signal events of 
both $0l$ and $1l$ type are sizable at  $\mathcal L =$ 
 1 $\ifb$ and their ratio - free from theoretical uncertainties - may
indicate the relic density producing mechanism.  
For example, at the point D (see figure 2 b) of \cite{debottam}) with
$m_0 = 80$, $m_{1/2} = 300$, $A_0 = -1000$, $tan \beta = 5$,
${({S \over \sqrt{B}})}_{1l} \approx 7$ {\footnote {At this point the  
predicted Higgs mass is little smaller than the LEP bound. However, due 
to the uncertainty of about 3 $GeV$ \cite{higgs} in the predicted $m_h$ this point is 
acceptable.}}. 
For  $m_0 = 100$, $m_{1/2} = 220$, $A_0 = -700$, $tan \beta = 10$ 
(point E; see fig 4 of \cite{debottam} ),
${({S \over \sqrt{B}})}_{1l} \approx 5$ 
. At the last two points bulk annihilation is the dominant generator of 
relic density. Finally at the point F 
$m_0 = 150$, $m_{1/2} = 200$, $A_0 = -600$, $tan \beta = 20$ (not studied
in \cite{debottam}) the $ {({S \over \sqrt{B}})}_{1l}  \approx 6$ for $ 
\mathcal L $ of 1 $\ifb$. As illustrated by  the last point 
the Higgs mass bound \cite{higgsbound}  can also be satisfied
for low $m_0$ and $m_{1/2}$ if $tan \beta$ is large even if $A_0$
is smaller than the values in A or B. Many of these points are also 
consistent with the relic
density constraint (see Fig. 3 of \cite{debottam} with $tan \beta$ = 30). 
The allowed points 
yield observable $0l$ signal but $1l$ signal is difficult to observe.    
This is due to the fact that most of the final states 
arising from electroweak gaugino decays contain $\tau$'s rather than 
$e$ and/or $\mu$.
For example, with
$m_0 = 170$, $m_{1/2} = 370$, $A_0 = -100$, $tan \beta = 30$
the ${({S \over \sqrt{B}})}_{0l} \approx 6$ for $ \mathcal L $ of 1 
$\ifb$. However, the $1l$ signal is unobservable.

It bears recall that the number of final states with tagged $\tau$-jets 
differs dramatically in the scenarios A, B and C 
\cite{debottam,nabanita}. We, therefore, turn our attention to final 
states 
of the type $1\tau + X$, where $X$
includes two or more hard jets but no $e$ or $\mu$ or tagged $\tau$. 
 
In ref. \cite{nabanita} $\tau$-jets  with
$P^{\tau-jet}_T >30 ~GeV$ and $\vert \eta \vert < 3.0 $
were shown to be taggable  according to the
efficiencies quoted by the CMS collaboration for different $P_T$ bins 
(see Fig. 12.9 of \cite{cms}). For $P_T \geq 130~GeV$ the 
tagging efficiency was as high as 0.90 while
for softer jets  the efficiencies were considerably smaller
(e.g., 0.50 for  $30~GeV \leq P_T \leq 50~GeV $).    
However. for $7~ TeV$ runs such information is
not available. As a reasonable guess 
we have assumed the overall $\tau$-jet tagging efficiency  
to be 
50$\%$ for $P^{\tau-jet}_T > 30 ~GeV$ and $\vert \eta \vert < 3.0$.  
The number of  $1\tau + X$ events   
for $\lum = 1 ~\ifb$ for signal A,  B and C turn  out to be 10, 5 and 
0.12 respectively. The other cuts are as stated above.
The total number of background events from the  sources in Table 4 is 
6.  The background from QCD arises due to mistagging of light jets.
A mistagging probability of 3$\%$ has been assumed for jets with $P_T > 
30 ~ GeV$.
%({\bf does this bkgd arise mainly from mistagging of QCD jets? if yes write a 
%line on the procedure}) . 
Considering this efficiency for 
scenario A the significance for $1\tau + X$  becomes 4.1
for $\lum = 1 ~\ifb$. For scenarios B and C the signal has poor 
significance. 
Thus if $\tau$-jet tagging can be implemented more efficiently and/ or 
the accumulated 
luminosity is larger than  $\lum = 1 ~\ifb$, then  $1\tau + X$ events may
provide another handle for discriminating different relic density 
producing scenarios.  

In this paper we have worked with leading order(LO) cross section only. This 
is because
the next to leading order (NLO) results for many of the backgrounds at 7 - 
$TeV$ are not 
known. From \cite{NLO} it is well known that the K 
factors for
the signal events vary from 1.3 - 1.4 for different signal processes at 14 -
$TeV$. In the third paper of \cite{7tev} the K-factor of squark-gluino 
production cross section was estimated to be approximately 1.3. Thus 
even if 
the NLO correction enhances the total background by a factor of two, the 
significances 
of the signals studied in this paper are not likely to change dramatically. 

%----------------------------------------------------------------
There are other mSUGRA scenarios consistent with relatively
light squark gluinos, the observed dark matter relic density
and the Higgs mass bound from LEP,
which could be of interest for the low energy runs of
the LHC. If one considers large $tan \beta$ ( $ \gsim$ 45)
the `Higgs funnel'  opens up 
\cite{funnel}.
%( Baer-Tata DM rev: ref 44)
Here a LSP  pair annihilates 
into  $b-\bar b$ through the $A$ (the pseudo
scalar Higgs) resonance to produce the observed relic density.
Sometimes the  $H$ (the heavy scalar Higgs) resonance also contributes.

A part of the
funnel region corresponds to light squarks and gluinos. The
observability of this region for $tan \beta =$ 45 at LHC-7 $TeV$ has been 
studied 
%(see  fig 2 of the third paper of ref 3 (Baer 7 TeV) . 
(see  Fig. 2 of the third paper of \cite{7tev}). 
It follows that if 
one takes into account the uncertainties in the theoretical prediction 
of the Higgs mass (i.e., parameter spaces with
111 $< m_h < $ 114 $GeV$ are considered to be allowed) a small  domain of 
the parameter 
space consistent with WMAP data can be probed with 1 $\ifb$ of
integrated luminosity.

A variation of the above theme is  the "$h$-pole region", where the 
condition
$2m_{\lspone} \lsim m_h$
is satisfied \cite{pnath,djouadi}. 
Here $\lspone$ annihilates through the exchange of a nearly on-shell
light CP even Higgs boson $h$. We present some example from \cite
{djouadi}  within the framework of mSUGRA. Here
large ranges of $m_0$ and $A_0$ are allowed but 
low $m_{1/2}$ is essential.
This leads to strong upper bounds on $m_{\chonep}$, $m_{\lsptwo}$
and most importantly on $m_{\tilde g}$ .
Hence the scenario could be interesting for
low energy runs.
Some examples from \cite{djouadi} are \\
(i)$m_t =  178 ~GeV$, $m_0 = 1500 ~GeV$, $A_0 = -1000 ~GeV$,
$tan \beta = 30$ leading to $m_h \sim 117 ~GeV$.\\
(ii)$m_t =  182 ~GeV$, $m_0 = 1000 ~GeV$, $A_0 = -1000 ~GeV$,
$tan \beta = 10$ leading to $m_h \sim 115 ~GeV$.\\
(iii)$m_t =  185 ~GeV$, $m_0 = 1000 ~GeV$, $A_0 = 0 ~GeV$,
$tan \beta = 20$ leading to $m_h \sim 116 ~GeV$.\\
In all the above cases DM allowed regions are obtained for $m_{1/2} < 145 ~GeV$. 
%(take from djouadi paper at various tan beta and mt
%, refer to djouadi fig for low mt).
Unfortunately, as already noted in \cite{djouadi}, this  resonance 
condition is satisfied 
if the physical mass of the top quark turns out to be on the
higher side which was favoured by the then available data. 
Such high $m_t$ are, however,  disfavoured by the central value of the 
current data  $m_t \approx$ 173 $GeV$.

In the mSUGRA scenario the scalars and the gaugino masses
are strictly universal at the GUT scale ($M_G$).
These conditions severely restrict the mSUGRA parameter space
consistent with the observed DM relic density.
If departures from the strict universality  are allowed 
\cite{nonuniv},
new regions of the parameter space consistent with both
the relic density data and light squark and gluinos
open up and novel LHC signatures are predicted. Most of the above works 
are, 
however, in the context of LHC-14 $TeV$. 

 There are  so many models for non-universality that
we were forced to restrict ourselves, perhaps quite arbitrarily, to 
brief comments on two scenarios only.
%Here are some examples from \cite{nonuniv}:
%(i) M3 light LM3DM comment on non zero $A_0$.
A dark matter allowed region can be obtained 
for relatively low values of $M_3$ (see the first paper of 
\cite{nonuniv})
due to non-universal gaugino masses at $M_G$.
For example, for
$m_0 = 300 ~GeV$, $M_1 = M_2 = 300 ~GeV$, $M_3 = 160 ~GeV$,
$A_0 = 0$, $tan \beta = 10$, $\mu > 1$ the value of $\Omega h^2 = 0.10$.
Although in this region $m_h =$ 106 $GeV$, which is disfavoured in spite of 
the uncertainty in $m_h$ noted above.

We have already argued that for negative values of $A_0$, consistency 
with the $m_h$ bound from LEP can be restored. Keeping all the parameters 
in the
last paragraph fixed and taking 
$A_0 = -600 ~GeV$ we obtain  $\Omega h^2 = 
0.12$. The  masses of
the sparticles (in $GeV$) of interest are:\\
$h          = 112$, 
$\tilde g   = 417$,
$\tilde u_L = 494$,
$\tilde u_R = 464$,
$\lstop     = 133$,
$\tilde b_1 = 406$,\\
$\tilde e_L = 364$,
$\tilde e_R = 322$,
$\stau_1    = 312$,
$\chonep    = 216$,
$\lsptwo    = 217$,
$\lspone    = 119$.

The above spectrum and the size of the signal estimated by us 
in this paper certainly 
suggest that an observable signal even at LHC-7 $TeV$ is a distinct 
possibility.

There is an economical way of introducing non-universality of
squark masses at $M_G$ by introducing a single parameter -
the $SO(10)$ D-term (for reference to earlier works and novel
signatures at Tevatron Run II see \cite{ad1,ad2}.
%AD et al hep-ph 9907444,
%A. Datta, A. Datta, M. Drees and D. P. Roy, \PRD(61,055003,2000).
%
%AD et al hep-ph 0007230)
%A. Datta, S. Maity and A. Datta, \JPG(27,1547,2001).
Here a possible scenario is lighter
down squarks of the R-type. These squarks along with a light gluino
can indeed lead to novel signals at LHC -7 $TeV$. 

%----------------------------------------------------------------

  We have deliberately refrained from imposing indirect constraints
on the mSUGRA parameter space, since these constraints invariably involve 
additional theoretical assumptions. The relaxation of such assumptions 
may drastically change the indirect constraints without affecting the
collider signatures and the relic density calculation. 

For example, the requirement that no charge colour breaking (CCB) breaking
minima of the scalar potential be deeper than the EWSB vacuum
(the unbounded from below (UFB)-3 constraint, the second paper of \cite{ccb}), 
puts lower bounds on sparticle 
masses \cite{avijit} which are stronger than the direct bounds from LEP in the 
mSUGRA model. However, such constraints loose their relevance if the 
EWSB minimum is a false vacuum with a life time larger than the age of the 
universe \cite{langaker}.
% ref 31 of debottam

Similarly the $ BR (b \rightarrow s \gamma)$ measured by the 
BABAR, BELLE and CLEO collaborations, is often used to
constrain the underlying theory.  However, 
the theoretical predictions have their share of uncertainties. 
The current data as quoted by
the Heavy Flavour Averaging Group (HEFAG) is 
$(3.52 \pm 23 \pm 9)\times 10^{-6}$ \cite{hfag}.
%0911.0217 ref 2
The improved  SM prediction (NLO) \cite{BSG} 
$(3.15 \pm 0.23 )\times 10^{-4}$,%0911.0217 ref 1
 though consistent with the data within
errors, leaves
ample room for a larger positive contribution from SUSY compared to the 
earlier estimates. This opens up the possibility of 
lighter sparticles in the mSUGRA model \cite{prannath}. 
%0911.0217
There are 
additional uncertainties as well. The assumption of minimal flavour violation,
which is
employed in theoretical computations, is not foolproof either. The relaxation 
of 
this assumption drastically weakens the constraints \cite{okumura}. 
%ref 30 of debottam
Even within the minimal flavour violation the inclusion 
of the CP violating phases  leads to further
uncertainties in the theoretical prediction (see, e.g., the second 
paper of \cite{okumura} and references there in).
It should be borne in mind that relaxation of the above assumptions
will have  very little or no impact on the direct collider signatures
considered by us.  

\section{The Signals at the LHC at 10 $TeV$}

In this section we briefly study the generic SUSY signals of the 
type
$m$-$l$ + $n$-$j$ + $\etslash $. Our aim is to study the
feasibility of discriminating among the three models in Table 1, should 
the performance of the 7 - $TeV$ run suggest yet another experiment at an 
energy less than the maximum attainable energy. To be specific we have 
considered a run at 10 $TeV$ which until recently was the favoured 
option.
%Next we shall employ flavour tagging and
%demonstrate that it further enhances our discriminatory power.
The total lowest order squark-gluino production 
cross-sections have been computed by CalcHEP \cite{calchep} and 
given in Table 5.
We have used the same selection criterion for jets and leptons as in
\cite{nabanita} for LHC-14 $TeV$. 
For background rejection we have used the CMS cuts which are also used in \cite{nabanita}.

%%%%%%%%%%%%%%%%%%%%%%%%%%%%%%%%%%%%%%%%%%%%%%%%%%%%%%%%%%%%%%%%%
%%%%%%%%%%%%%%%%%%%%%%%%%%%%%%%%%%%%%%%%%%%%%%%%%%%
%\begin{table}[!ht]
%\begin{center}\

%\begin{tabular}{|c|c|c|c|}
%\hline
%     & \multicolumn{3}{c|}{}
%& \multicolumn{3}{c|}{$\sigma(\pb)$}\\
%\cline{2-4}
%Process&A&B&C\\
%\hline
%$\wt g \wt g$            &   0.293  & 0.098 & 0.0054\\
%$\qL \wt g$              &   0.720  & 0.293 & 0.0274\\
%$\qR \wt g$              &   0.780 & 0.325 & 0.0311\\
%$\qL \qL$                &   0.318  & 0.168 & 0.0301\\
%$\qL \qL^*$              &   0.108  & 0.029 & 0.0023\\
%$\qR \qR$                &   0.283  & 0.150 & 0.0276\\
%$\qR \qR^*$              &   0.095  & 0.037 & 0.0032\\
%$\qL \qR$                &   0.205  & 0.099 & 0.0144\\
%$\qL \qR^* + c.c$        &   0.149  & 0.067 & 0.0077\\
%$\lstoppair$             &   2.08  & 0.365 & 0.0037\\
%$\hstoppair$             &   0.02  & 0.008 & 0.0006\\
%$\lstop \hstopbar + c.c$ &   0.0002 & 0.0002 &6.8$\times 10^{-6}$\\
%$\lsbotpair$             &   0.047  & 0.017 &0.0008\\
%$\hsbotpair$             &   0.019  & 0.008 &0.0006\\
%\hline
%Total &5.12 &1.663 &0.1548 \\
%\hline

%\end{tabular}
%\end{center}
%   \caption{The production cross sections of all squark-gluino events
%studied in this paper.}
%\end{table}

%%%%%%%%%%%%%%%%%%%%%%%%%%%%%%%%%%%%%%%%%%%%%%%%%%%%%%%%%%%%%%%%%

%%%%%%%%%%%%%%%%%%%%%%%%%%%%%%%%%%%%%%%%%%%%%%%%%%%

\begin{table}[!ht]
\begin{center}\
\begin{tabular}{|c|c|c|c|}
%\hline
%&\multicolumn{3}{c|}{SIGNAL} & \multicolumn{3}{c|}{background}\\
%\cline{2-6}
\hline
& A &B&C  \\
\hline
$\sigma (\pb)$     & 5.12   & 1.663   & 0.1548   \\
\hline \hline
$ 0 l $ & 0.5628 & 0.3238 & 0.0434  \\
\hline
$ 1 l $ & 0.0488 & 0.0393 & 0.0186  \\
\hline
$ S S $ &0.00072  & 0.000698 & 0.00134 \\
\hline
$ O S $ & 0.00154  & 0.002278 & 0.00359\\
\hline
$ 3l $ & 0.000051  & .000133 & .00066 \\
\hline
\end{tabular}
\end{center}
   \caption{The cross-sections (including efficiency) at $Q = \sqrt{\hat s}  $ for signal process
with different $m$. Here $SS$ refers to $m = 2$ with leptons carrying the
same charge and $OS$ refers to similar events with leptons carrying opposite charge.}
\label{sig10}
\end{table}
%%%%%%%%%%%%%%%%%%%%%%%%%%%%%%%%%%%%%%%%%%%%
\begin{table}[!ht]
\begin{center}\
\begin{tabular}{|c|c|c|c|c|c|c|c|}
%\hline
%&\multicolumn{3}{c|}{SIGNAL} & \multicolumn{3}{c|}{background}\\
%\cline{2-6}
\hline
                  & $t\bar t$ & $W + 1j$ & $W + 2j$& $Z+ 1j$&$Z+2j$ & QCD&Total \\
\hline
$\sigma (\pb)$    & 170       & 1549.5   &  468.1  & 577.9  & 127.6   & 758& Background\\
\hline \hline
$ 0 l $           & 0.0657    & 0.0638   & 0.1379  & 0.0364 & 0.1089 &0.904&1.3167 \\
\hline
$ 1 l $           & 0.0315    & 0.0550   & 0.0944  & 0.00173 & 0.0021&0.0015&0.1862 \\
\hline
$ S S $           & 0.00017    & -   & -  & - & -&-&0.00017 \\
\hline
$ O S $           & 0.00323    & -   & -  & - & -&-&0.00323 \\
\hline

\end{tabular}
\end{center}
   \caption{The cross-sections (including efficiency) at $Q = \sqrt{\hat s}  $ for background process
with different $m$. $3 l$ is background free.
%Here $SS$ refers to $m = 2$ with leptons carrying the
%same charge and $OS$ refers to similar events with leptons carrying opposite charge.
No entry in a particular column (-) means negligible background.
}
\label{bgnd10}
\end{table}

%%%%%%%%%%%%%%%%%%%%%%%%%%%%%%%%%%%%%%%%%%%%%%%%%%%%%%%%%%%%%%%%%
\begin{table}
\begin{center}\
\begin{tabular}{|c|c|c|c|}
\hline
 & A & B & C \\
%\cline{2-6}
\hline
$ 0 l $ &49.1 & 28.2  & 3.8  \\
\hline
$ 1 l $ & 11.3  & 9.1  & 4.3  \\
\hline
$  SS $ &  5.5  & 5.4  & 10.3  \\
\hline
$  OS $ &  2.7  &  4.0  & 6.3  \\
\hline
\end{tabular}
\end{center}
   \caption{The significance (S/$\sqrt B$)of signals in Table \ref{sig10} for
$ \mathcal L$$ = 10 ~\ifb $.}
\label{significance}
\end{table}
%************
\begin{table}[!ht]
\begin{center}\
\begin{tabular}{|c|c|c|c|}
%\hline
%&\multicolumn{3}{c|}{SIGNAL} &\multicolumn{3}{c|}{BACKGROUNDS}\\
%\cline{2-7}
\hline
& A &B&C \\
\hline
$\sigma (\pb)$     & 5.12   & 1.663   & 0.1548   \\

\hline $1 \tau + X$& 0.0663   & 0.0337  & 0.0042\\
\hline
$1 e  + X$         & 0.012   & 0.0096  & 0.0057 \\
\hline
\end{tabular}
\end{center}
   \caption{The cross-sections (including efficiency) of events with one detected $\tau$
and one isolated $e$. Here $X$ stands for all possible final states excluding any lepton
or tagged $\tau$ but with at least two jets.
The  number of tagged $b$-jets is given by $n$-$b$, $n=1,2,3$ (see the first
column  of row 4 - 9)
.}
\label{lepsig}
\end{table}

%%%%%%%%%%%%%%%%%%%%%%%%%%%%%%%%%%%%%%%%%%%%%%%%%%%%%%%%%%%%%%%%%
\begin{table}[!ht]
\begin{center}\
\begin{tabular}{|c|c|c|c|c|c|c|c|}
\hline
                  & $t\bar t$ & $W + 1j$ & $W + 2j$& $Z+ 1j$&$Z+2j$ & QCD&Total \\
\hline
$\sigma (\pb)$    & 170       & 1549.5   &  468.1  & 577.9  & 127.6   & 758& Background\\

\hline $1 \tau + X$& 0.0104   & 0.0087   & 0.0236  & 0.00116 & 0.00456  & 0.142&0.1904\\
\hline
$1 e  + X$         & 0.0095   & 0.0203   & 0.0309  & 0.0006  & 0.00035  & 0.0003&0.0620\\
\hline
\end{tabular}
\end{center}
   \caption{Same as Table \ref{lepsig} for backgrounds.}
%The cross-sections (including efficiency) of events with one detected $\tau$
%and one isolated $e$. Here $X$ stands for all possible final states excluding any lepton
%or tagged $\tau$ but with at least two jets.
%The  number of tagged $b$-jets is given by $n$-$b$, $n=1,2,3$ (see the first
%column  of row 4 - 9)
%.}
\label{lepbgnd}
\end{table}

%%%%%%%%%%%%%%%%%%%%%%%%%%%%%%%%%%%%%%%%%%%%%%%%%%%%%%%%%%%%%%%%%
\begin{table}[!ht]
\begin{center}\
\begin{tabular}{|c|c|c|c|}
\hline
%&\multicolumn{3}{c|}{SIGNAL}\\
%\cline{2-4}
%\hline
& A &B&C\\
\hline
%$\sigma (\pb)$ & 15.58& 5.74 & 0.74\\
%\hline
$1 \tau + X$       & 15.2  & 7.7  & 0.9\\
\hline
$1 e + X$          & 4.8  & 3.6  & 2.3\\
\hline
\end{tabular}
\end{center}
   \caption{ The S/$\sqrt B$ ratio for the signals in Table \ref{lepsig} 
   corresponding to $ \mathcal L$$ = 10 ~\ifb $. }
  \label{lepsignificance} 
\end{table}

%%%%%%%%%%%%%%%%%%%%%%%%%%%%%%%%%%%%%%%%%%%%%%%%%%%%%%%%%%%%%%%%%

In Table \ref{sig10} we have shown a  multi-channel analysis using signals with
different choices of $m$ (the number of leptons in the final state) which
efficiently discriminate among different scenarios.
We present
in Table \ref{bgnd10} the important standard model backgrounds
in the leading order for Q = $\sqrt {\hat s}$.
%For $SS$ and $OS$ the major background stems from $t \bar t$ (see
%table 6).
%The raw cross -sections for $SS$ and $OS$ are 0.00017 and 0.00323
%respectively. 
The $t \bar t$ and $QCD$ backgrounds have been calculated in the same way
as in \cite{nabanita}. 
For $W$ + $n$-$jets$ and $Z$ + $n$-$jets$ events we have generated 
events with $n=0,1$ and $2$ at the parton level using ALPGEN (v 2.13) 
\cite{alpgen}. We have generated these events with $P_T^j > 80~GeV$. 
These partonic events have been fed to Pythia for parton showering,
hadronozation, fragmentation and decays etc.

%**** dicuss about W+ jets and z+ jets background.

In Table \ref{significance} we have presented the significance ($
S \over {\sqrt B}$), where $ S(B)$ is the total number of
signal (background) events for integrated luminosity  
$ \mathcal L $ of 10 $\ifb$.

Although the squark-gluino production cross-section is rather tiny in
scenario C the signal cross-sections predicted for $m \geq 2$ is larger than
the corresponding signals in A and B with much larger raw production 
cross-section.
This again is a direct consequence of the large  BRs of
electroweak gaugino decays into $e$ and $\mu$.

From Table \ref{significance}  it is clear that one way to unambiguously 
discriminate  between A, B on the one hand and C on the other,
is the count of $0l$ and $1l$ events and their ratio free from
theoretical ambiguities. These signals are not visible
in scenario C. In contrast the visibility of $SS$ and $OS$
signals in scenario C is much better. However, 
the observation of the almost
background free signal for $m = 3$ is the best bet for establishing  C
since no statistically significant signal is expected from A or B in 
this case, although this may require
$\lum \geq  20 ~\ifb$.

Obviously the number of final states involving
$\tau$ leptons, a critical observable for discriminating among 
the models, cannot be established  without invoking $\tau$-jet tagging 
in the analysis.
We, therefore, turn our attention to final states of the type $1\tau + X$
where $X$
includes two or more hard jets but no $e$ or $\mu$ or tagged $\tau$. Tagging of
$\tau$-jets are implemented as in \cite{nabanita} which followed the 
CMS analysis.

The computation of  $1e + X$ type events are rather straight forward.
 Here for simplicity we have assumed the
e-detection efficiency to be 100 \%. In our generator level analysis the
result for $1\mu + X$ is expected to be the same to a good approximation
and we do not present them separately. It should be borne in mind that
in this case harder cuts on e has been implemented to exclude the 
electrons from the leptonic $\tau$ decays. 
The QCD background
to  $1\tau + X$ events stems
from mistagging of light flavour jets as $\tau$-jets.
The mistagging probability has also been taken from \cite{cms} Fig. 
12.9. Table \ref{lepsig} contains the  $1 \tau + X$ and $1e + X$  signals.
The dominant SM backgrounds have been listed in Table \ref{lepbgnd}.

The statistical significance of various
 signals for the representative value $\cal L $ of 10 $\ifb$ are
 listed in Table \ref{lepsignificance}.
 The  $1\tau + X$  signal, if unambiguously observed,
 will disfavour model C. 

\section{Conclusions}

The mSUGRA parameter space with relatively low $m_0$ and $\mhalf$ corresponding
to light squarks and gluinos is consistent with the WMAP data on DM 
relic density and the lower bound on the lightest Higgs scalar mass from LEP, 
provided the trilinear coupling $A_0$ has large to moderate negative 
values \cite{debottam}. Scanning the parameter space by generating 
squark-gluino events with the event generator Pythia we find that
the jets + $\etslash$ signal is observable over a significant parameter 
space (see Tables 3 and 4 and the discussions in section 3) 
for an integrated luminosity of $\lum = 1 ~\ifb$. In a smaller but 
non-trivial parameter space the 1-$l$ + jets + $\etslash$ signal is also 
observable. The probability of observing the latter increases
if a little larger integrated luminosity $\lum = (2-3) ~\ifb$ is available. 
If both the signals are observable then the ratio of the number of events 
which is remarkably free from theoretical uncertainties may help 
distinguishing among different scenarios \cite{nabanita}. However, 
the parameter space with the conventional choice $A_0 = 0$ 
where LSP - $\stau_1$ coannihilation is the only relic density generating 
mechanism does not yield any observable signal. 

As already noted in \cite{debottam} the $1\tau + X$ signal, where $X$
includes two or more hard jets but no $e$ or $\mu$ or tagged $\tau$ is
a very good discriminator for different mechanisms for relic density 
production. In the absence of any information about the $\tau$-jet tagging 
efficiency at 7 - $TeV$ from simulations by the experimentalists, 
we have assumed an overall efficiency  
of 50$\%$ for $P^{\tau-jet}_T > 30 ~GeV$ and $\vert \eta \vert< 3.0$. 
With this choice no observable signal emerge for $\lum = 1 ~\ifb$. 
However, if $\tau$-jet tagging efficiency and/or $\lum$ increases, then an 
additional handle for DM signatures at 7 - $TeV$ will be available.

Since the performance of the 7 - $TeV$ run may dictate yet another 
round of experiments at an energy less than the maximum attainable energy, 
we have considered the above signal at LHC-10 $TeV$. Here a full multichannel 
analysis of  generic SUSY signals consisting of $m$-leptons + $n$-jets + 
$\etslash$  without any flavour tagging can probe the parameter space
in great details (see Tables  5 - 7). Of course if efficient flavour tagging is 
available a more powerful discriminator of different relic density producing 
mechanisms will be in operation (see Tables 8 - 10).

%----------------------------------------------------------\

{\bf Acknowledgment}:
NB
would like to thank the Council of Scientific and
Industrial Research, Govt. of India for financial support.

%----------------------------------------------------------
%%%%%%%%%%%%%%%%%%%%%%%%%%%%%%%%%%%%%%%%%%%%%%%%%%%%
%\newpage

\end{document}

%% file: definition.tex
\topmargin -0.1in
\headsep 30pt
\footskip 40pt
\oddsidemargin 12pt
\evensidemargin -16pt
\textheight 8.5in
\textwidth 6.5in
\parindent 20pt
 
\def\baselinestretch{1.5}
%%\def\baselinestretch{1.1}
%-------- General character and macros start -----------------
\newcommand{\newc}{\newcommand}
\def\preprint{{preprint}}
\def\Ord{\lower .7ex\hbox{$\;\stackrel{\textstyle <}{\sim}\;$}}
\def\OOrd{\lower .7ex\hbox{$\;\stackrel{\textstyle >}{\sim}\;$}}
% --- calligraphic stuff ---
\def\cO#1{{\cal{O}}\left(#1\right)}
%\DeclareRobustCommand*\cal{\@fontswitch\relax\mathcal}
\newc{\order}{{\cal O}}
\def\lag             {{\cal L}}
\def\Lag             {{\cal L}}
\def\lum             {{\cal L}}
\def\R               {{\cal R}}
\def\Rsq             {{\cal R}^{\sq}}
\def\Rst             {{\cal R}^{\st}}
\def\Rsb             {{\cal R}^{\sb}}
\def\M               {{\cal M}}
\def\Oas             {{\cal O}(\alpha_{s})}
\def\Vcal            {{\cal V}}
\def\Wcal            {{\cal W}}
\newc{\be}{\begin{equation}}
\newc{\ee}{\end{equation}}
\newc{\br}{\begin{eqnarray}}
\newc{\er}{\end{eqnarray}}
\newc{\ba}{\begin{array}}
\newc{\ea}{\end{array}}
\newc{\bi}{\begin{itemize}}
\newc{\ei}{\end{itemize}}
\newc{\bn}{\begin{enumerate}}
\newc{\en}{\end{enumerate}}
\newc{\bc}{\begin{center}}
\newc{\ec}{\end{center}}
\newc{\ul}{\underline}
\newc{\ol}{\overline}
\newc{\ra}{\rightarrow}
%\newc{\to}{\ra}
\newc{\lra}{\longrightarrow}
\newc{\wt}{\widetilde}
\newc{\til}{\tilde}
\def\kr              {^{\dagger}}
\newc{\wh}{\widehat}
\newc{\ti}{\times}
\newc{\Dir}{\kern -6.4pt\Big{/}}
\newc{\Dirin}{\kern -10.4pt\Big{/}\kern 4.4pt}
\newc{\DDir}{\kern -10.6pt\Big{/}}
\newc{\DGir}{\kern -6.0pt\Big{/}}
\newc{\sig}{\sigma}
\newc{\sigmalstop}{\sig_{\lstoppair}}
\newc{\Sig}{\Sigma}  %%\def\Sig{\Sigma} this is the def style
\newc{\del}{\delta}
\newc{\Del}{\Delta}
\newc{\lam}{\lambda}
\newc{\Lam}{\Lambda}
\newc{\gam}{\gamma}
\newc{\Gam}{\Gamma}
\newc{\eps}{\epsilon}
\newc{\Eps}{\Epsilon}
\newc{\kap}{\kappa}
\newc{\Kap}{\Kappa}
\newc{\modulus}[1]{\left| #1 \right|}
\newc{\eq}[1]{(\ref{eq:#1})}
\newc{\eqs}[2]{(\ref{eq:#1},\ref{eq:#2})}
\newc{\etal}{{\it et al.}\ }
\newc{\ibid}{{\it ibid}.}
\newc{\ibidem}{{\it ibidem}.}
\newc{\eg}{{\it e.g.}\ }
\newc{\ie}{{\it i.e.}\ }
\def \viz{\emph{viz.}}
\def \etc{\emph{etc. }}
\newc{\nonum}{\nonumber}
\newc{\lab}[1]{\label{eq:#1}}
\newc{\dpr}[2]{({#1}\cdot{#2})}
\newc{\lt}{\stackrel{<}}
\newc{\gt}{\stackrel{>}}
\newc{\lsimeq}{\stackrel{<}{\sim}}
\newc{\gsimeq}{\stackrel{>}{\sim}}
\def\lsim{\buildrel{\scriptscriptstyle <}\over{\scriptscriptstyle\sim}}
\def\gsim{\buildrel{\scriptscriptstyle >}\over{\scriptscriptstyle\sim}}
%\newcommand{\lsim}{\raisebox{-0.13cm}{~\shortstack{$<$ \\[-0.07cm] $\sim$}}~}
%\newcommand{\gsim}{\raisebox{-0.13cm}{~\shortstack{$>$ \\[-0.07cm] $\sim$}}~}
%sim single and approx double sim
\def\lapp{\mathrel{\rlap{\raise.5ex\hbox{$<$}}
                    {\lower.5ex\hbox{$\sim$}}}}
\def\gapp{\mathrel{\rlap{\raise.5ex\hbox{$>$}}
                    {\lower.5ex\hbox{$\sim$}}}}
\newc{\half}{\frac{1}{2}}
\newcommand {\nnc}        {{\overline{\mathrm N}_{95}}}
\newcommand {\dm}         {\Delta m}
\newcommand {\dM}         {\Delta M}
\def\bra{\langle}
\def\ket{\rangle}
\def\cO#1{{\cal{O}}\left(#1\right)}
\def \DM{{\Delta{m}}}
\newc{\bQ}{\ol{Q}}
\newc{\dota}{\dot{\alpha }}
\newc{\dotb}{\dot{\beta }}
\newc{\dotd}{\dot{\delta }}
\newc{\nindnt}{\noindent}

% Def. for small \fracs:
\newcommand{\medf}[2] {{\footnotesize{\frac{#1}{#2}} }}
\newcommand{\smaf}[2] {{\textstyle \frac{#1}{#2} }}
\def\onesq            {{\textstyle \frac{1}{\sqrt{2}} }}
\def\onehf            {{\textstyle \frac{1}{2} }}
\def\oneth            {{\textstyle \frac{1}{3} }}
\def\twoth            {{\textstyle \frac{2}{3} }}
\def\onefo            {{\textstyle \frac{1}{4} }}
\def\forth            {{\textstyle \frac{4}{3} }}

\newc{\matth}{\mathsurround=0pt}
\def\ML{\ifmmode{{\mathaccent"7E M}_L}
             \else{${\mathaccent"7E M}_L$}\fi}
\def\MR{\ifmmode{{\mathaccent"7E M}_R}
             \else{${\mathaccent"7E M}_R$}\fi}
\newcommand{\s}{\\ \vspace*{-3mm} }

\def \ud { {1 \over 2} }
\def \ut { {1 \over 3} }
\def \td { {3 \over 2} }
\newc{\mr}{\mathrm}
\def\dh {\partial }
\def \cs { cross-section }
\def \css { cross-sections }
\def \cm { centre of mass }
\def \cms { centre of mass energy }
\def \cc { coupling constant }
\def \ccs {coupling constants }
\def \gc {gauge coupling }
\def \gcc {gauge coupling constant }
\def \gccs {gauge coupling constants }
\def \yc {Yukawa coupling }
\def \ycc {Yukawa coupling constant }
\def \pp {{parameter }}
\def \pps {{parameters }} % spdas {{ or { which is correct 
% if not take from alephstop
\def \ps {parameter space }
\def \pss {parameter spaces }
\def \vv {vice versa }

\newc{\siminf}{\mbox{$_{\sim}$ {\small {\hspace{-1.em}{$<$}}}    }}
\newc{\simsup}{\mbox{$_{\sim}$ {\small {\hspace{-1.em}{$>$}}}    }}

%------------------------------------------------------------

% SM notations
\newc {\Zboson}{{\mathrm Z}^{0}}
\newc{\thetaw}{\theta_W}
\newc{\mbot}{{m_b}}
\newc{\mtop}{{m_t}}
\newc{\sm}{${\cal {SM}}$}
\newc{\as}{\alpha_s}
\newc{\aem}{\alpha_{em}}
\def \PI{{\pi^{\pm}}}
\newc{\ppbar}{\mbox{$p\ol{p}$}}
\newc{\bbbar}{\mbox{$b\ol{b}$}}
\newc{\ccbar}{\mbox{$c\ol{c}$}}
\newc{\ttbar}{\mbox{$t\ol{t}$}}
\newc{\eebar}{\mbox{$e\ol{e}$}}
\newc{\zzero}{\mbox{$Z^0$}}
\def \gamz{\Gam_Z}
\newc{\wplus}{\mbox{$W^+$}}
\newc{\wminus}{\mbox{$W^-$}}
\newc{\ellp}{\ell^+}
\newc{\ellm}{\ell^-}
\newc{\elp}{\mbox{$e^+$}}
\newc{\elm}{\mbox{$e^-$}}
\newc{\elpm}{\mbox{$e^{\pm}$}}
\newc{\qbar}     {\mbox{$\ol{q}$}}
%%\newc{\lp}{\mbox{$e^+$}}
%%\newc{\lm}{\mbox{$e^-$}}
%%\newc{\lpm}{\mbox{$e^{\pm}$}}
\def \ewgroup{SU(2)_L \otimes U(1)_Y}
\def \smgroup{SU(3)_C \otimes SU(2)_L \otimes U(1)_Y}
\def \smcolorem{SU(3)_C \otimes U(1)_{em}}
%---------------------------------------------------------

%SUSY notations
\def \SSM  {Supersymmetric Standard Model}
\def \poincare{Poincare$\acute{e}$}
\def \superspace{\emph{superspace}}
\def \sfs{\emph{superfields}}
\def \superpot{\emph{superpotential}}
\def \csf{\emph{chiral superfield}}
\def \csfs{\emph{chiral superfields}}
\def \vsf{\emph{vector superfield }}
\def \vsfs{\emph{vector superfields}}
\newc{\Ebar}{{\bar E}}
\newc{\Dbar}{{\bar D}}
\newc{\Ubar}{{\bar U}}
\newc{\susy}{{{SUSY}}}
\newc{\msusy}{{{M_{SUSY}}}}
%----------------------------------------------------

%Gauginos
\def\photino{\ifmmode{\mathaccent"7E \gam}\else{$\mathaccent"7E \gam$}\fi}
\def\taugluino{\ifmmode{\tau_{\mathaccent"7E g}}
             \else{$\tau_{\mathaccent"7E g}$}\fi}
\def\mphotino{\ifmmode{m_{\mathaccent"7E \gam}}
             \else{$m_{\mathaccent"7E \gam}$}\fi}
\newc{\gl}   {\mbox{$\wt{g}$}}
\newc{\mgl}  {\mbox{$m_{\gl}$}}
%%\newc{\gl}{\wt g}
%%\newc{\mgl}{m_{\gl}}
%%\def\gluino{\ifmmode{\mathaccent"7E g}\else{$\mathaccent"7E g$}\fi}
%%\def\mgluino{\ifmmode{m_{\mathaccent"7E g}}
%%             \else{$m_{\mathaccent"7E g}$}\fi}
%----------------------------------------------------
% Chargino
\def \charginopm{{\wt\chi}^{\pm}}
\def \mcharginopm{m_{\charginopm}}
\def \mchpmmin {\mcharginopm^{min}}
\def \chonep {{\wt\chi_1^+}}
\def \chone {{\wt\chi_1}}
\def \ch2p {{\wt\chi_2^+}}
\def \chonem {{\wt\chi_1^-}}
\def \ch2m {{\wt\chi_2^-}}
\def \chplus {{\wt\chi^+}}
\def \chminus {{\wt\chi^-}}
\def \chonip{{\wt\chi_i}^{+}}
\def \chonim{{\wt\chi_i}^{-}}
\def \chonipm{{\wt\chi_i}^{\pm}}
\def \chonjp{{\wt\chi_j}^{+}}
\def \chonjm{{\wt\chi_j}^{-}}
\def \chonjpm{{\wt\chi_j}^{\pm}}
\def \chonepm{{\wt\chi_1}^{\pm}}
\def \chonemp{{\wt\chi_1}^{\mp}}
\def \mchonepm{m_{\chonepm}}
\def \mchonemp{m_{\chonemp}}
\def \chtwopm{{\wt\chi_2}^{\pm}}
\def \mchtwopm{m_{\chtwopm}}
\newc{\dmchi}{\Delta m_{\wt\chi}}

%-----------------------------------------------------------------------
% Neutralino

\def \vlsp{\emph{VLSP}}
\def \lspi{\wt\chi_i^0}
\def \mlspi{m_{\lspi}}
\def \lspj{\wt\chi_j^0}
\def \mlspj{m_{\lspj}}
\def \lspone{\wt\chi_1^0}
\def \mlspone{m_{\lspone}}
\def \lsptwo{\wt\chi_2^0}
\def \mlsptwo{m_{\lsptwo}}
\def \lspthree{\wt\chi_3^0}
\def \mlspthree{m_{\lspthree}}
\def \lspfour{\wt\chi_4^0}
\def \mlspfour{m_{\lspfour}}

%-----------------------------------------------------------------------
%SLEPTONs

\newc{\sele}{\wt{\mathrm e}}
\newc{\sell}{\wt{\ell}}
\def \msell{m_{\sell}}
\def \slepone{\wt\ell_1}
\def \mslepone{m_{\slepone}}
\def \smuone{\wt\mu_1}
\def \msmuone{m_{\smuone}}
\def \stauone{\wt\tau_1}
\def \mstauone{m_{\stauone}}
\def \snu{\wt{\nu}}
\def \snutau{\wt{\nu}_{\tau}}
\def \msnu{m_{\snu}}
\def \msnumu{m_{\snu_{\mu}}}
\def \barsnu{\wt{\bar{\nu}}}
\def \barsnul{\barsnu_{\ell}}
\def \snul{\snu_{\ell}}
\def \mbarsnu{m_{\barsnu}}
\newc{\snue}     {\mbox{$ \wt{\nu_e}$}}
\newc{\smu}{\wt{\mu}}
\newc{\stau}{\wt{\tau}}
%%% from 9810232.tex
\newc {\nuL} {\wt{\nu}_L}
\newc {\nuR} {\wt{\nu}_R}
\newc {\snub} {\bar{\wt{\nu}}}
\newc {\eL} {\wt{e}_L}
\newc {\eR} {\wt{e}_R}
%%% from 9810232.tex
\def \slepl{\wt{l}_L}
\def \mslepl{m_{\slepl}}
\def \slepr{\wt{l}_R}
\def \mslepr{m_{\slepr}}
\def \stau{\wt\tau}
\def \mstau{m_{\stau}}
\def \slepton{\wt\ell}
\def \mslepton{m_{\slepton}}
\def \mlhiggs{m_{h^0}}

%----------------------------------------------------
%SQUARKs
\def \xr{X_{r}}

%\newc{\sq}   {\mbox{$\wt{q}$}}
%\newc{\msq}  {\mbox{$m_(\sq)$}}
\def \sfer{\wt{f}}
\def \msfer{m_{\sfer}}
\def \sq{\wt{q}}
\def \msq{m_{\sq}}
\def \msquleft{m_{\tilde{u_L}}}
\def \msqurht{m_{\tilde{u_R}}}
\def \sql{\wt{q}_L}
\def \msql{m_{\sql}}
\def \sqr{\wt{q}_R}
\def \msqr{m_{\sqr}}
\newc{\msqot}  {\mbox{$m_(\sq_{1,2} )$}}
\newc{\sqbar}    {\mbox{$\bar{\wt{q}}$}}
\newc{\ssb}      {\mbox{$\squark\ol{\squark}$}}
\newc {\qL} {\wt{q}_L}
\newc {\qR} {\wt{q}_R}
\newc {\uL} {\wt{u}_L}
\newc {\uR} {\wt{u}_R}
\def \ul{\wt{u}_L}
\def \mul{m_{\ul}}
\newc {\dL} {\wt{d}_L}
\newc {\dR} {\wt{d}_R}
\newc {\cL} {\wt{c}_L}
\newc {\cR} {\wt{c}_R}
\newc {\sL} {\wt{s}_L}
\newc {\sR} {\wt{s}_R}
\newc {\tL} {\wt{t}_L}
\newc {\tR} {\wt{t}_R}
\newc {\stb} {\ol{\wt{t}}_1}
\newc {\sbot} {\wt{b}_1}
\newc {\msbot} {m_{\sbot}}
\newc {\sbotb} {\ol{\wt{b}}_1}
\newc {\bL} {\wt{b}_L}
\newc {\bR} {\wt{b}_R}
\def \mul{m_{\wt{u}_L}}
\def \mur{m_{\wt{u}_R}}
\def \mdl{m_{\wt{d}_L}}
\def \mdr{m_{\wt{d}_R}}
\def \mcl{m_{\wt{c}_L}}
\def \charml{\wt{c}_L}
\def \mcr{m_{\wt{c}_R}}
\newc{\csquark}  {\mbox{$\wt{c}$}}
\newc{\csquarkl} {\mbox{$\wt{c}_L$}}
\newc{\mcsl}     {\mbox{$m(\csquarkl)$}}
\def \msl{m_{\wt{s}_L}}
\def \msr{m_{\wt{s}_R}}
\def \mbl{m_{\wt{b}_L}}
\def \mbr{m_{\wt{b}_R}}
\def \mtl{m_{\wt{t}_L}}
\def \mtr{m_{\wt{t}_R}}
\def \st{\wt{t}}
\def \mst{m_{\st}}
\newc {\stopl}         {\wt{\mathrm{t}}_{\mathrm L}}
\newc {\stopr}         {\wt{\mathrm{t}}_{\mathrm R}}
\newc {\stoppair}      {\wt{\mathrm{t}}_{1}
\bar{\wt{\mathrm{t}}}_{1}}
\def \lstop{\wt{t}_{1}}
\def \lstopbar{\lstop^*}
\def \hstop{\wt{t}_{2}}
\def \hstopbar{\hstop^*}
\def \mlstop{m_{\lstop}}
\def \mhstop{m_{\hstop}}
\def \lstoppair{\lstop\lstop^*}
\def \hstoppair{\hstop\hstop^*}
\newc{\tsquark}  {\mbox{$\wt{t}$}}
\newc{\ttb}      {\mbox{$\tsquark\ol{\tsquark}$}}
\newc{\ttbone}   {\mbox{$\tsquark_1\ol{\tsquark}_1$}}
% top squark related
\def \tsq {top squark }
\def \tsqs {top squarks }
\def \tsql {ligtest top squark }
\def \tsqh {heaviest top squark }
\newc{\mix}{\theta_{\wt t}}
\newc{\cost}{\cos{\theta_{\wt t}}}
\newc{\sint}{\sin{\theta_{\wt t}}}
\newc{\costloop}{\cos{\theta_{\wt t_{loop}}}}
%-----------------------------------------------------------------------
\def \lsbot{\wt{b}_{1}}
\def \lsbotbar{\lsbot^*}
\def \hsbot{\wt{b}_{2}}
\def \hsbotbar{\hsbot^*}
\def \mlsbot{m_{\lsbot}}
\def \mhsbot{m_{\hsbot}}
\def \lsbotpair{\lsbot\lsbot^*}
\def \hsbotpair{\hsbot\hsbot^*}
\newc{\mixsbot}{\theta_{\wt b}}

%-----------------------------------------------------------------------
%HIGGS sector
\def \mhone{m_{h_1}}
\def \hup{{H_u}}
\def \hdn{{H_d}}
\newc{\tb}{\tan\beta}
\newc{\cb}{\cot\beta}
\newc{\vev}[1]{{\left\langle #1\right\rangle}}

%-----------------------------------------------------------------------
%SOFT BRAEAKING TERMS
\def \abot{A_{b}}
\def \atop{A_{t}}
\def \atau{A_{\tau}}
\newc{\mhalf}{m_{1/2}}
\newc{\mzero} {\mbox{$m_0$}}
\newc{\azero} {\mbox{$A_0$}}

%--------------------------------------------------------------
% RPV related  stuff
\newc{\lb}{\lam}
\newc{\lbp}{\lam^{\prime}}
\newc{\lbpp}{\lam^{\prime\prime}}
\newc{\rpv}{{\not \!\! R_p}}
\newc{\rpvm}{{\not  R_p}}
%%%\newc{\rpv}{\not R_p}
%%\newc{\rpv}{{\not \!\! R_p}}
\newc{\rp}{R_{p}}
%\newc{\rpmssm}{{$R_p$-MSSM}\ }
%\newc{\rpvmssm}{$\rpv$-MSSM}
\newc{\rpmssm}{{RPC MSSM}}
\newc{\rpvmssm}{{RPV MSSM}}

%---------------------------------------------------------------
% Collider related stuffs

\newc{\sbyb}{S/$\sqrt B$}
\newc{\pelp}{\mbox{$e^+$}}
\newc{\pelm}{\mbox{$e^-$}}
\newc{\pelpm}{\mbox{$e^{\pm}$}}
\newc{\epem}{\mbox{$e^+e^-$}}
\newc{\lplm}{\mbox{$\ell^+\ell^-$}}
\def \branch{\emph{BR}}
\def \branche{\branch(\lstop\ra be^{+}\nu_e \lspone)\ti \branch(\lstop^{*}\ra \bar{b}q\bar{q^{\prime}}\lspone)}
\def \branchmu{\branch(\lstop\ra b\mu^{+}\nu_{\mu} \lspone)\ti \branch(\lstop^{*}\ra \bar{b}q\bar{q^{\prime}}\lspone)}
\def\Ecm{\ifmmode{E_{\mathrm{cm}}}\else{$E_{\mathrm{cm}}$}\fi}
\newc{\rts}{\sqrt{s}}
\newc{\rtshat}{\sqrt{\hat s}}
\newc{\gev}{\,GeV}
\newc{\mev}{~{\rm MeV}}
\newc{\tev}  {\mbox{$\;{\rm TeV}$}}
\newc{\gevc} {\mbox{$\;{\rm GeV}/c$}}
\newc{\gevcc}{\mbox{$\;{\rm GeV}/c^2$}}
\newc{\intlum}{\mbox{${ \int {\cal L} \; dt}$}}
\newc{\call}{{\cal L}}
%%\def \miset{\not\!\!{E_T}}
%%\newc{\etmiss}{/ \hskip-7pt E_T}
%%\def \mispt{p{\!\!\!/}_T} 
\def \met  {\mbox{${E\!\!\!\!/_T}$}}
\def \cpv  {\mbox{${CP\!\!\!\!/}$}}
\newc{\ptmiss}{/ \hskip-7pt p_T}
\def \eslash{\not \! E}
\def \etslash{\not \! E_T }
\def \ptslash{\not \! p_T }
\newc{\PT}{\mbox{$p_T$}}
\newc{\ET}{\mbox{$E_T$}}
\newc{\dedx}{\mbox{${\rm d}E/{\rm d}x$}}
\newc{\ifb}{\mbox{${\rm fb}^{-1}$}}
\newc{\ipb}{\mbox{${\rm pb}^{-1}$}}
\newc{\pb}{~{\rm pb}}
\newc{\fb}{~{\rm fb}}
\newc{\ycut}{y_{\mathrm{cut}}}
\newc{\chis}{\mbox{$\chi^{2}$}}
\def \hadron{\emph{hadron}}
\def \nlc{\emph{NLC }}
\def \lhc{\emph{LHC }}
\def \cdf{\emph{CDF }}
\def\dzero{\emptyset}
\def \tevatron{\emph{Tevatron }}
\def \lep{\emph{LEP }}
\def \jets{\emph{jets }}
\def \jet(s){\emph{jet(s) }}

%-----------------------------------------------------------------------
%  Light Stop decay parameters and different modes:
\def\Crs{stroke [] 0 setdash exch hpt sub exch vpt add hpt2 vpt2 neg V currentpoint stroke 
hpt2 neg 0 R hpt2 vpt2 V stroke}
\def\loopdk{\lstop \ra c \lspone}
\def\brloopdk{\branch(\loopdk)}
\def\fourdk{\lstop \ra b \lspone  f \bar f'}
\def\brfourdk{\branch(\fourdk)}
\def\fourdklep{\lstop \ra b \lspone  \ell \nu_{\ell}}
\def\fourdkhad{\lstop \ra b \lspone  q \bar q'}
\def\brfourdklep{\branch(\fourdklep)}
\def\brfourdkhad{\branch(\fourdkhad)}
\def\twodk{\lstop \ra b \chonep}
\def\brtwodk{\branch(\twodk)}
\def\threedkslep{\lstop \ra b \wt{\ell} \nu_{\ell}}
\def\brthreedkslep{\branch(\threedkslep)}
\def\threedksnu{\lstop \ra b \wt{\nu_{\ell}} \ell }
\def\brthreedksnu{\branch(\threedksnu) }
\def\threedklsp{\lstop \ra b W \lspone }
\def\brthreedklsp{\\branch(\threedklsp) }
\def\topdk{t \ra \lstop \lspone}
\def\rpvdk{\lstop \ra e^+ d}
\def\brrpvdk{\branch(\rpvdk)}
\def\fonec{f_{11c}} 
%-----------------------------------------------------------------------
%  Scale  of Physics
\newc{\mpl}{M_{\rm Pl}}
\newc{\mgut}{M_{GUT}}
\newc{\mw}{M_{W}}
\newc{\mweak}{M_{weak}}
\newc{\mz}{M_{Z}}

%--Collabortaions ----------------------------------------
\newc{\OPALColl}   {OPAL Collaboration}
\newc{\ALEPHColl}  {ALEPH Collaboration}
\newc{\DELPHIColl} {DELPHI Collaboration}
\newc{\XLColl}     {L3 Collaboration}
\newc{\JADEColl}   {JADE Collaboration}
\newc{\CDFColl}    {CDF Collaboration}
\newc{\DXColl}     {D0 Collaboration}
\newc{\HXColl}     {H1 Collaboration}
\newc{\ZEUSColl}   {ZEUS Collaboration}
\newc{\LEPColl}    {LEP Collaboration}
\newc{\ATLASColl}  {ATLAS Collaboration}
\newc{\CMSColl}    {CMS Collaboration}
\newc{\UAColl}    {UA Collaboration}
\newc{\KAMLANDColl}{KamLAND Collaboration}
\newc{\IMBColl}    {IMB Collaboration}
\newc{\KAMIOColl}  {Kamiokande Collaboration}
\newc{\SKAMIOColl} {Super-Kamiokande Collaboration}
\newc{\SUDANTColl} {Soudan-2 Collaboration}
\newc{\MACROColl}  {MACRO Collaboration}
\newc{\GALLEXColl} {GALLEX Collaboration}
\newc{\GNOColl}    {GNO Collaboration}
\newc{\SAGEColl}  {SAGE Collaboration}
\newc{\SNOColl}  {SNO Collaboration}
\newc{\CHOOZColl}  {CHOOZ Collaboration}
\newc{\PDGColl}  {Particle Data Group Collaboration}

%--- Macros of the journal start----------------------------------------
\def\issue(#1,#2,#3){{\bf #1}, #2 (#3)}%AIP format!Vol,page(Year)
% In thesis Journal macros is exhibits to be Vol, (Year), Page
% and in the bibliography it is set like
% (vol,page,Year ) and then the below macros  
% will take care everything.
%\def\issue(#1,#2,#3){{\bf #1} (#3) #2 } % PLB format!Vol,(Year),page
\def\ASTR(#1,#2,#3){Astropart.\ Phys. \issue(#1,#2,#3)}
\def\AJ(#1,#2,#3){Astrophysical.\ Jour. \issue(#1,#2,#3)}
\def\AJS(#1,#2,#3){Astrophys.\ J.\ Suppl. \issue(#1,#2,#3)}
\def\APP(#1,#2,#3){Acta.\ Phys.\ Pol. \issue(#1,#2,#3)}
\def\JCAP(#1,#2,#3){Journal\ XX. \issue(#1,#2,#3)} %spdas
\def\SC(#1,#2,#3){Science \issue(#1,#2,#3)}
\def\PRD(#1,#2,#3){Phys.\ Rev.\ D \issue(#1,#2,#3)}
\def\PR(#1,#2,#3){Phys.\ Rev.\ \issue(#1,#2,#3)} % spdas check 
\def\PRC(#1,#2,#3){Phys.\ Rev.\ C \issue(#1,#2,#3)}
\def\NPB(#1,#2,#3){Nucl.\ Phys.\ B \issue(#1,#2,#3)}
\def\NPPS(#1,#2,#3){Nucl.\ Phys.\ Proc. \ Suppl \issue(#1,#2,#3)}
\def\NJP(#1,#2,#3){New.\ J.\ Phys. \issue(#1,#2,#3)}
\def\JP(#1,#2,#3){J.\ Phys.\issue(#1,#2,#3)}
\def\PL(#1,#2,#3){Phys.\ Lett. \issue(#1,#2,#3)}
\def\PLB(#1,#2,#3){Phys.\ Lett.\ B  \issue(#1,#2,#3)}
\def\ZP(#1,#2,#3){Z.\ Phys. \issue(#1,#2,#3)}
\def\ZPC(#1,#2,#3){Z.\ Phys.\ C  \issue(#1,#2,#3)}
\def\PREP(#1,#2,#3){Phys.\ Rep. \issue(#1,#2,#3)}
\def\PRL(#1,#2,#3){Phys.\ Rev.\ Lett. \issue(#1,#2,#3)}
\def\MPL(#1,#2,#3){Mod.\ Phys.\ Lett. \issue(#1,#2,#3)}
\def\RMP(#1,#2,#3){Rev.\ Mod.\ Phys. \issue(#1,#2,#3)}
\def\SJNP(#1,#2,#3){Sov.\ J.\ Nucl.\ Phys. \issue(#1,#2,#3)}
\def\CPC(#1,#2,#3){Comp.\ Phys.\ Comm. \issue(#1,#2,#3)}
\def\IJMPA(#1,#2,#3){Int.\ J.\ Mod. \ Phys.\ A \issue(#1,#2,#3)}
\def\MPLA(#1,#2,#3){Mod.\ Phys.\ Lett.\ A \issue(#1,#2,#3)}
\def\PTP(#1,#2,#3){Prog.\ Theor.\ Phys. \issue(#1,#2,#3)}
\def\RMP(#1,#2,#3){Rev.\ Mod.\ Phys. \issue(#1,#2,#3)}
\def\NIMA(#1,#2,#3){Nucl.\ Instrum.\ Methods \ A \issue(#1,#2,#3)}
\def\JHEP(#1,#2,#3){J.\ High\ Energy\ Phys. \issue(#1,#2,#3)}
\def\JPG(#1,#2,#3){J.\ Phys. \ G \issue(#1,#2,#3)}
\def\EPJC(#1,#2,#3){Eur.\ Phys.\ J.\ C \issue(#1,#2,#3)}
\def\RPP (#1,#2,#3){Rept.\ Prog.\ Phys. \issue(#1,#2,#3)}
\def\PPNP(#1,#2,#3){ Prog.\ Part.\ Nucl.\ Phys. \issue(#1,#2,#3)}
\newc{\PRDR}[3]{{Phys. Rev. D} {\bf #1}, Rapid  Communications, #2 (#3)}
%%%%%%%%%%%%%%%%%%%%%%%%%%%%%%%%%%%%%%%%%%%%%%%%%%%%%%%%%%%%%%%